\documentclass[a4paper,11pt]{article}

\usepackage{jcappub} 

\usepackage{amsmath}
	\usepackage{graphicx}
	\usepackage{makeidx}
	\usepackage{amsfonts}
	\usepackage{gensymb}
	\usepackage[ansinew]{inputenc}
	\usepackage[usenames,dvipsnames]{pstricks}
	\usepackage{subfigure}
	\usepackage{epsfig}
	\usepackage{pst-grad} 
	\usepackage{pst-plot} 

\title{\boldmath Comparing Galactic Center MSSM dark matter solutions to the Reticulum II gamma-ray data}

\author[a]{Abraham Achterberg}
\author[a]{Melissa van Beekveld}
\author[a,c]{Wim Beenakker}
\author[a,b]{Sascha Caron}
\author[a]{Luc Hendriks}

\affiliation[a]{Institute for Mathematics, Astrophysics and Particle Physics,\\
                Faculty of Science, Mailbox 79,\\
                Radboud University Nijmegen, P.O. Box 9010, NL-6500 GL Nijmegen,
                The Netherlands}
\affiliation[b]{Nikhef, Science Park,\\ Amsterdam, The Netherlands}
\affiliation[c]{Institute  of  Physics,  University  of  Amsterdam,\\  Science  Park  904,  1018  XE  Amsterdam,  The
Netherlands}

\emailAdd{a.achterberg@astro.ru.nl}
\emailAdd{mcbeekveld@gmail.com}
\emailAdd{wimb@hef.ru.nl}
\emailAdd{scaron@cern.ch}
\emailAdd{luc.hendriks@gmail.com}

\abstract{Observations with the Fermi Large Area Telescope (LAT) indicate a possible small photon signal originating from the dwarf galaxy Reticulum II that exceeds the expected background between 2 GeV and 10 GeV. We have investigated two specific
scenarios for annihilating WIMP dark matter within the phenomenological Minimal Supersymmetric Standard Model (pMSSM) framework as a possible source for these photons. We find that the same parameter ranges in pMSSM as reported by an earlier paper to be consistent with the Galactic Center excess, are also consistent with the excess observed in Reticulum II, resulting in a $J$-factor of $\log_{10}(J(\alpha_{int}=0.5\degree)) \simeq (20.3-20.5)^{+0.2}_{-0.3}$ GeV$^2$cm$^{-5}$. This $J$-factor is consistent with $\log_{10}(J(\alpha_{int}=0.5\degree)) = 19.6^{+1.0}_{-0.7}$  GeV$^2$cm$^{-5}$, which was derived using an optimized spherical Jeans analysis of kinematic data obtained from the Michigan/Magellan Fiber System (M2FS).}


\begin{document}
\maketitle
\flushbottom

\section{Introduction}
\noindent

\noindent While the existence of dark matter (DM) is widely accepted and its cosmological abundance is measured, the fundamental nature of DM remains unknown. The most promising explanation is that DM is a neutral and weakly interacting particle, not described by the Standard Model (SM). A weakly interacting massive particle (WIMP) is a leading scenario for dark matter, as it can lead to the right abundance of DM originating from a thermal freeze out in the early universe. WIMPs can be searched for with several detection strategies, one of which being indirect detection via annihilation products (for example gamma rays) \cite{Jungman:1995df, Bertone:2004pz}. \\

\noindent Observations of our Galactic Center (GC) with the Fermi-LAT satellite over the past several years, show that there is a significant excess in gamma rays in the energy range of $1 \; {\rm GeV} \lesssim E_{\gamma} \lesssim  50 \; {\rm GeV}$ (see refs. \cite{Goodenough:2009gk, Vitale:2009hr, Hooper:2010mq, Hooper:2011ti,Abazajian:2012pn, Gordon:2013vta, Macias:2013vya, Abazajian:2014fta,Daylan:2014rsa, Zhou:2014lva, Calore:2014xka, simonaTalk, Gaggero:2015nsa, Cholis:2015dea, O'Leary:2015gfa}  for details and interpretations). The source of these photons is under debate. Astrophysical explanations (like pulsars or cosmic ray outbursts) have been shown to face challenges, but it is not yet possible to rule out these explanations \cite{Cholis:2014lta, Hooper:2013nhl, Petrovic:2014xra, Petrovic:2014uda, Carlson:2014cwa, Bartels:2015aea, Lee:2015fea}. Another possibility for the origin of this $\gamma$-ray excess is annihilating DM. The GC is a promising target for searches of DM signals, as large-scale simulations of galaxy formation predict DM halos around galaxies such as the Milky way \cite{NFW:1997, Mill:2009}. This implies that an indirect DM signal could come from the Galactic Center \cite{Mo:2006}.\\

\noindent Initially, the GC excess signal was reported to be compatible with a 30 GeV dark matter particle annihilating into $b\bar{b}$, or a 10 GeV dark matter particle annihilating to $\tau \bar{\tau}$ \cite{Hooper:2010mq, 2011PhRvD..84l3005H}. The resulting $b$-quarks and $\tau$-leptons can hadronize to neutral pions, which can decay to the observed $\gamma$-ray photons. Both processes should have an annihilation cross section close to $\langle\sigma v\rangle \simeq 1.3 \times 10^{-26}$ cm$^3$s$^{-1}$, the annihilation cross section expected for a thermal relic with a mass at the weak scale $E_w \simeq 100$ GeV. These predictions have been explored within various contexts, for example within the Minimal Supersymmetric Standard Model (MSSM). In the framework of the MSSM, it is impossible to obtain such a scenario given LEP constraints \cite{Cahill-Rowley:2014ora}. \\
However, a study on astrophysical background model systematics \cite{Calore:2014xka} allows for also higher WIMP masses and different annihilation channels to give a good fit to the GC excess \cite{Agrawal:2014oha}. In addition it is shown that uncertainties originating from high energy physics (especially partron showers) are also important to take into account \cite{Caron:2015wda}. Recently new annihilation scenarios within the phenomenological MSSM (pMSSM) and compatible with all the constraints are found \cite{Caron:2015wda, Bertone:2015tza}. In the pMSSM the 105 Lagrangian parameters of the MSSM are reduced to 22, using phenomenological constraints on several parameters. In this model we assume that:
\begin{itemize}
\item Masses of the first and second generation sfermions are equal, separately in the lepton and quark sectors.
\item All the soft SUSY-breaking parameters are real, so the only source for CP-violation is the CKM matrix in the standard model.
\item Requiring minimal flavor violation, in the MSSM all sfermion mass matrices are assumed to be diagonal. 
\end{itemize}
As a result of R-parity conservation in the MSSM, a lightest supersymmetric particle (LSP) exists. In most scenarios the LSP is the lightest neutralino, $\tilde{\chi}^0_1$, which is a combination of the neutral electroweak gaugino and higgsino fields. We select only models with a neutralino as LSP. The neutralino is a well-motivated DM candidate, as it is neutral, stable and can lead to a right DM relic density \cite{vanBeekveld:2015tka}. A 19-dimensional realization of the pMSSM is enough to encapsulate the  phenomenology of the 22-parameter model. This is achieved by putting all trilinear couplings of the first and second generation to 0. The remaining parameters are 10 sfermion masses, 3 gaugino masses $M_{1,2,3}$, the ratio of the Higgs vacuum expectation values $\tan\beta$, the Higgsino mixing parameter $\mu$, the mass $m_A$ of the CP-odd Higgs-boson $A^0$ and 3 trilinear couplings $A_{b,t,\tau}$.  \\

\noindent Dwarf galaxies (dSphs) in the vicinity of our galaxy provide an alternative to the Galactic Center for the search for $\gamma$-rays originating from annihilating dark matter. The amount of DM in these objects can be estimated by measuring the stellar velocities of member stars \cite{Bonnivard:2015xpq}. The resulting DM signal is proportional to the line-of-sight integral of the DM density distribution, a quantity known as the $J$-factor.  Dwarf galaxies have a lower DM density (typical up to $\log_{10}(J) = 20$ GeV$^2$ cm$^{-5}$) than expected for the Galactic Center (for which the $J$-factor is predicted to be at least an order of magnitude larger). However, dSphs have less complicated backgrounds than the Galactic Center, as the latter suffers from systematic uncertainties associated with diffuse fore- and backgrounds \cite{2013PhR...531....1S}. These facts make dSphs promising targets for the detection of $\gamma$-rays originating from DM annihilation.  Until recently, no dSphs with a significant $\gamma$-ray excess were found, setting strong limits on the cross section for DM annihilation. Recent imaging data from the Dark Energy Survey (DES) led to the discovery of new Milky Way satellites \cite{Abbott:2005bi, 2015ApJ...805..130K, 2015ApJ...807...50B}. Among those, Reticulum II is of particular interest, as this object showed a gamma ray excess of a global significance level of 2.3 - 3.7 $\sigma$ \cite{Hooper:2015ula, Geringer-Sameth:2015lua} between 2 GeV and 10 GeV. The Fermi and DES collaborations found a less significant excess (p-value of 0.05 including a trials factor from testing multiple DM masses and channels, corresponding to a global significance of 1.65 $\sigma$), using an updated (Pass 8) data set \cite{Drlica-Wagner:2015xua}. The local significance they find is 2.4 $\sigma$ (p = 0.01).\\ 

\noindent In this paper we will present a comparison of the Pass 7 data for Reticulum II in terms of annihilating neutralino dark matter. We use the pMSSM as a framework to provide this dark matter particle, using the results of ref. \cite{Caron:2015wda}. This study has no channel and mass trials factor since the DM mass and shape are fixed by the pMSSM model and the Galactic Center excess data. We will update this study to Pass 8 data in due course. 

\newpage
\section{Analysis set-up}

\begin{figure}[b]
	\begin{center}
		\includegraphics[width=0.7\textwidth]{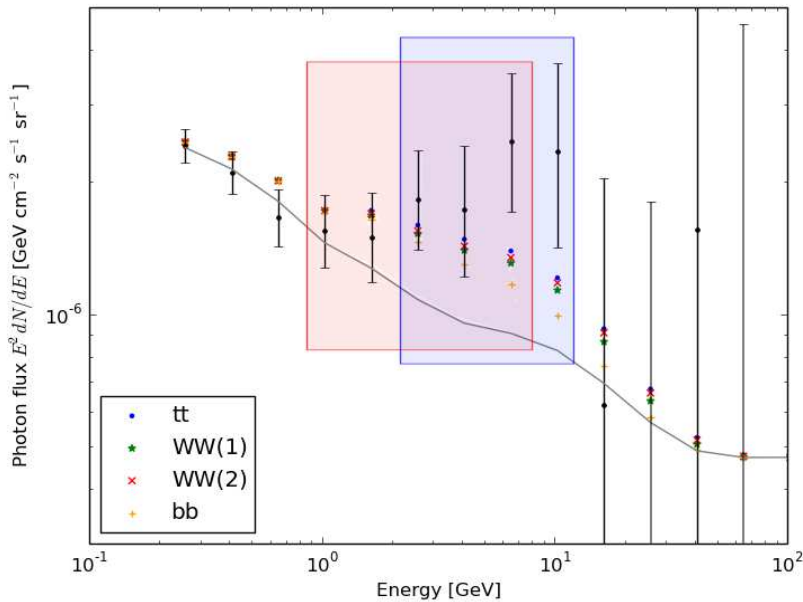}
		\caption{Photon spectrum as extracted from ref. \cite{Geringer-Sameth:2015lua} of the Fermi-LAT data using events within 0.5$\degree$ of Reticulum II (black points) with Poisson error bars. The solid gray line is the sum of background estimates of the Fermi Collaboration's models for isotropic and Galactic diffuse emission at the location of Reticulum II. Between 2 GeV and 10 GeV, the photon spectrum rises above the background. This observed excess region is indicated by the blue box. The colored points are fluxes generated by DarkSUSY of the best fits for four scenarios: tt (blue), WW(1) (green), WW(2) (red) as extracted from \cite{Caron:2015wda} and bb (orange), added to the background of Reticulum II. The red box indicates the expected excess region, as determined by equation \ref{eq:maxi}.} 
		\label{fig:reticulum}
	\end{center}
\end{figure}

\noindent In ref. \cite{Geringer-Sameth:2015lua} events are included within 0.5$\degree$ from Reticulum II. We adopt the same strategy and will refer to this as the region of interest (ROI). See figure \ref{fig:reticulum} for the resulting photon spectrum, which is extracted from ref. \cite{Geringer-Sameth:2015lua}. The error bars indicate Poisson confidence level intervals (68$\%$) on the number of counts in each bin. A logarithmic binning of 5 bins per decade between 0.2 and 300 GeV is used, resulting in a total of 15 bins. In this figure the last 2 bins are not shown, because of their large uncertainties. It is clear that between 2 and 10 GeV, the spectrum of Reticulum II rises above the expected background. We will refer to this region as the observed excess region from now on. See ref. \cite{Geringer-Sameth:2015lua} for details on the analysis leading to this gamma-ray excess. From the ROI we derive the integration angle to be $\alpha_{int} = 0.5\degree$, corresponding to $\Omega = 2\pi(1-\cos{\alpha_{int}}) \simeq 2.4 \times 10^{-4}$ sr. \\

\noindent The observed gamma-ray flux from DM annihilation per unit solid angle at a photon energy $E_{\gamma}$ is given by:
\begin{equation}
    \frac{{\rm d}\Phi_{\gamma}(E_{\gamma})}{{\rm d}E_{\gamma}{\rm d} \Omega} =
    \frac{\langle \sigma v \rangle}{8\pi m_{\rm DM}^2}\frac{{\rm d}N_\gamma}{{\rm d}E}
    \int_{\text{l.o.s.}} \: \rho_{\rm DM}^2(l) {\rm d}l\;,
\end{equation}
where the integral over the DM density squared is along the line of sight (l.o.s.), $\langle\sigma v\rangle$ is the 
annihilation cross section weighted by the relative velocity, $m_{\rm DM}$ is the WIMP mass, and $dN_{\gamma}/dE$ denotes the photon spectrum per annihilation. \\
The normalization factor $\int {\rm d}\Omega \int_{\text{l.o.s.}}\: \rho_{\rm DM}^2(l){\rm d}l\;$ is commonly referred to as the $J$-factor:
\begin{equation}
\label{eq:J}
J = \int {\rm d}\Omega \int_{\text{l.o.s.}} \rho_{\rm DM}^2(l) {\rm d} l
\end{equation}

\noindent We use SUSPECT \cite{Djouadi:2002ze} as a spectrum generator and DarkSUSY 5.1.1 \cite{Gondolo:2004sc, DarkSUSY} to compute the photon fluxes. In ref. \cite{Caron:2015wda}, it is shown that there are three pMSSM parameter ranges that are consistent with the GC photon excess as originating from annihilating DM. That paper contains the details on how the possible solutions were found using a scan of the pMSSM parameter space. The solutions presented in ref. \cite{Caron:2015wda} yield a value for the DM relic density that corresponds to $\Omega_{\rm DM} h^2 \simeq 0.1$, close to the observed value $\Omega_{\rm DM} h^2 = 0.12\pm0.0027$ \cite{Planck:2015}, without being constrained a-priori to do so. In all cases the lightest neutralino is the DM candidate. 

\noindent The first region of the pMSSM parameter space that was found to yield a good fit to the GC excess is one in which the lightest neutralinos annihilate mostly to a pair of top quarks. The lightest neutralinos have a very dominant bino component (99$\%$), and their masses are at the kinematical threshold ($m_{\chi} \sim 174 - 187$ GeV). These models have $\Omega_{\rm DM} h^2 = 0.08 - 0.21$. We will refer to these models as tt. \\
The second solution is $\tilde{\chi}^0_1\tilde{\chi}^0_1 \rightarrow W^+W^-$. In this case the lightest neutralino is a bino-higgsino mixture, with a mass range of $m_{\chi} \sim 84 - 92$ GeV. We will refer to these models as WW(1). This solution provided the best fit to the GC excess (p-value=0.45). These models correspond to a $\Omega_{\rm DM} h^2$ in the range of 0.08-0.10.\\
The third solution yields a scenario in which the lightest neutralinos annihilate mostly to $W^+W^-$ pairs. In this case, the composition of the lightest neutralinos is a mixture of bino, wino and higgsino, where the bino component is the most dominant one (90\%). The mass of the lightest neutralinos is $m_{\chi} \sim 87 - 97$ GeV. The Galactic Center best fit points have an $\Omega_{\rm DM} h^2$ of 0.07 - 0.18. We will refer to these models as WW(2).\\
\noindent In table \ref{tab:ranges} we show the exact parameter ranges in the pMSSM for each scenario. From each scenario we adopt the 50 best fit models (corresponding with the best $\chi^2$ for the Galactic Center excess fit) for comparison with the Reticulum II signal. \\

\begin{table}[b]
\begin{center}
\begin{tabular}{|c|c|c|c|c|}
\hline
& $M_1$ (GeV) & $M_2$ (GeV) & $\mu$ (GeV) & $\tan\beta$ \\
\hline
tt &$171 - 189$ & $ 190-1550 $ & $> 250$ & $> 5$ \\
WW(1) &$103 - 119$ & $ 240-660 $ & $108-142$ & $8-50$ \\
WW(2) &$91 - 101$ & $ 102-127 $ & $156-507$ & $5-12$ \\
\hline
\end{tabular}

\caption{Parameter ranges in pMSSM that correspond to the best solutions to the GC excess, as extracted from \cite{Caron:2015wda}.}
\label{tab:ranges}
\end{center}
\end{table}
\noindent In addition to these three scenarios, we create a 50 GeV neutralino and consider only the $b\bar{b}$ annihilation channel. We artificially put $\langle\sigma v\rangle \simeq 1.3 \times 10^{-26}$ cm$^3$s$^{-1}$, the upper limit for the annihilation cross section as found by ref. \cite{Ackermann:2015zua} and used in ref. \cite{Hooper:2015ula} to explain the Reticulum II excess. We will refer to this model as bb. 
\newpage

\section{Results}


\noindent The resulting photon fluxes are fitted with the photon spectrum of Reticulum II, leaving the $J$-factor as a fitting parameter. Since the error bars are approximately symmetric and uncorrelated, we use the following $\chi^2$ definition to determine the best fit:
\begin{equation}
\label{eq:chi}
\chi^2 = \sum_i \frac{(d_i - m_i)^2}{\sigma_i^2},
\end{equation}
where $i$ is the energy bin number, running from 1 to 15, $d_i$ is the Reticulum flux and $m_i$ the DarkSUSY model flux. In figure \ref{fig:reticulum} we show the resulting photon spectrum for the best fits. We use (minimal $\chi^2)+1$ to define the 1$\sigma$ range of the $J$-factor. In the following, we give the mean values of this $J$-factor error for all 50 best fit models within each scenario. \\
\begin{figure}[b]
	\begin{center}
		\includegraphics[width=0.7\textwidth]{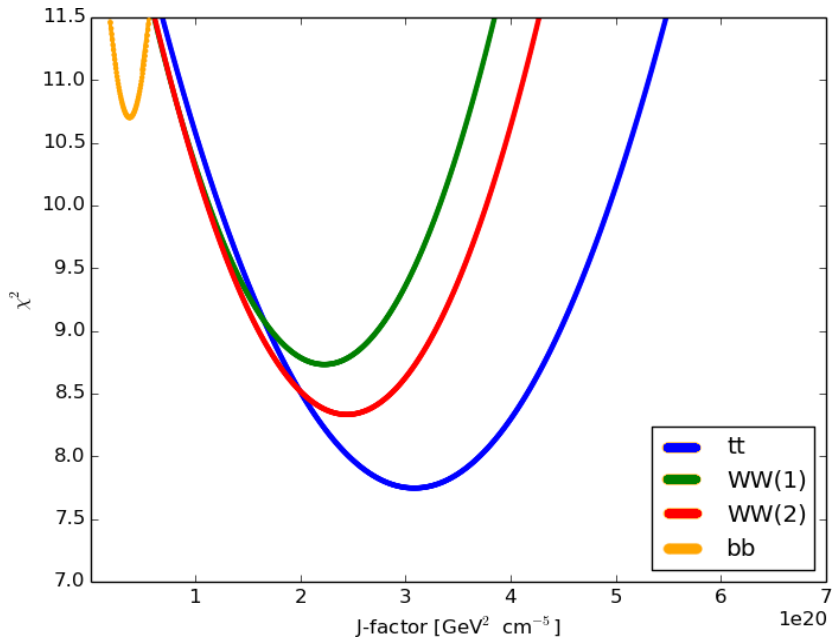}
		\caption{$J$-factor (note we do not use the logarithm) of the three best fits within tt (blue), WW(1) (green), WW(2) (red) and bb (orange) as a function of $\chi^2$ (as defined in equation \ref{eq:chi}).} 
		\label{fig:chisq}
	\end{center}
\end{figure}

\noindent We find that the tt scenarios yield the best results, with a minimal $\chi^2$ ranging between 7.75 and 7.93. To determine the corresponding p-value we take a conservative approach and use 13 degrees of freedom (d.o.f.). The d.o.f. are counted as follows: there is a total of 15 bins, which gives 15 degrees of freedom. We subtract 1 d.o.f. for the background fit and 1 d.o.f. for the $J$-factor normalization. Using 13 d.o.f., the chi2-values correspond to a p-value of $\simeq 0.85$. The tt models give a $J$-factor of $\log_{10}(J(\alpha_{int})) = (20.33 - 20.56) ^{+0.15}_{-0.23}$ GeV$^2$cm$^{-5}$. The numbers between brackets denote the range of the optimal fit $J$-factor for each model. \\

\noindent The WW(1) scenario results in a $J$-factor of $\log_{10}(J(\alpha_{int})) = (20.31 - 20.35) ^{+0.16}_{-0.25}$ GeV$^2$cm$^{-5}$. The chi2-values are slightly worse with a mean value of 8.75, corresponding to a p-value of 0.79 using 13 d.o.f. \\
\noindent The WW(2) scenario gives a $J$-factor of $\log_{10}(J(\alpha_{int})) = (20.25 - 20.55) ^{+0.15}_{-0.25}$ GeV$^2$cm$^{-5}$. These have a $\chi^2$ between 8.35 and 8.66, corresponding to a p-value of 0.81. \\
The last scenario, bb, results in $\log_{10}(J(\alpha_{int})) = 19.57^{+0.19}_{-0.35}$ if we use an annihilation cross section of $\langle\sigma v\rangle \simeq 1.3 \times 10^{-26}$ cm$^3$s$^{-1}$. The bb scenario has $\chi^2= 10.7$, which corresponds to a p-value of 0.63. \\
In figure \ref{fig:chisq} we show the $J$-factor as a function of $\chi^2$ for one model in each scenario. Note that we do not plot the logarithm of the $J$-factor in this figure. \\

\noindent In table \ref{tab:pvalue} we summarize the best p-values of the total signal and the p-values for the observed excess region for each scenario (indicated by \emph{Reticulum II data} and \emph{Observed excess region}). In order to provide for an unbiased statistical test of the signal shapes, we also include p-values for an expected excess region and for a background only model. We define the expected excess region as the range of bins where: 
\begin{equation}
\label{eq:maxi}
\min\left\{{p_i({\rm signal + background}|{\rm background})}\right\}
\end{equation}
is satisfied, where a minimum number of bins of 4 is demanded, because the observed excess region consists of 4 bins. The signal refers to the photon flux as generated by DarkSUSY. In equation \ref{eq:maxi}, the label $i$ denotes a certain continuous bin range and $p$ denotes the p-value. $\chi^2$ is calculated according to equation \ref{eq:chi} and the number of d.o.f. is equal to the number of bins - 1 (note that the $J$-factor normalization is now set, so we gain one extra d.o.f.). This range of bins corresponds to the situation where the best discrepancy between signal as generated by DarkSUSY and background can be made. We adopt the relative Poisson distributed errors on the Reticulum II measured data as an error on the background. The expected excess region is for all four models the same and indicated by the red box in figure~\ref{fig:reticulum}. This region differs from the observed excess region by a few bins. \\
 
\noindent As shown in table \ref{tab:pvalue}, the tt scenario yields the best fit results for the observed excess region (p=0.36) and expected excess region (p=0.53). The bb scenario yields the worst fit (p=0.09 for the observed excess region, p=0.24 for the expected excess region). \\
\begin{table*}[b]
\begin{tabular}{|c|c|c|c|}
\hline
& Reticulum II data & Observed excess region & Expected excess region\\
\hline
tt & 0.85 & 0.36 & 0.53 \\
WW(1) & 0.79 &  0.27 & 0.36 \\
WW(2) & 0.81 & 0.31 & 0.40\\
bb & 0.63 & 0.09 & 0.24\\
background & 0.37 & 0.01 & 0.03\\
\hline
\end{tabular}
\caption{The p-values for the proposed four scenarios corresponding to the Reticulum II measured data, the observed excess region from 2 to 10 GeV (as indicated in figure \ref{fig:reticulum} by the blue box) and the expected excess region (as indicated in figure \ref{fig:reticulum} by the red box). In the last row we include p-values for the background only model to explain the Reticulum II measured data, the observed excess region and the expected excess region. } 
\label{tab:pvalue}
\end{table*} 

\noindent Our conclusion is that the DM annihilation models, as proposed in ref. \cite{Caron:2015wda} which are consistent with the gamma ray excess in the Galactic Center as reported by Fermi-LAT, provide also a good fit to the observed gamma ray excess in Reticulum II. We find here that all these models predict a $J$-factor between $\log_{10}(J(\alpha_{int})) = 20.0$ GeV$^2$cm$^{-5}$ and $\log_{10}(J(\alpha_{int})) = 20.7$ GeV$^2$cm$^{-5}$ (including $1\sigma$ error). The bb model results in a $J$-factor of $\log_{10}(J(\alpha_{int})) = 19.57^{+0.19}_{-0.35}$. The GC models lie within the 1$\sigma$ region of $\log_{10}(J(\alpha_{int})) = 19.6^{+1.0}_{-0.7}$  GeV$^2$cm$^{-5}$ for an integration angle of $\alpha_{int} = 0.5 \degree$ as reported by ref. \cite{Bonnivard:2015tta}.  This derivation is done using an optimized spherical Jeans analysis of kinematic data obtained from the Michigan/Magellan Fiber System, an analysis which is independent of our approach. Ref. \cite{2015arXiv150402889S} reported a lower $J$-factor of $\log_{10}(J(\alpha_{int}=0.5\degree)) = 18.9\pm0.6$ GeV$^2$cm$^{-5}$ (1$\sigma$ uncertainty), using the same spectroscopic data. This difference could originate from a different choice of dark matter priors (such as the dark matter local density and halo profile). Our models are consistent within 2$\sigma$ with this $J$-factor. \\


\section{Summary and conclusions}
\noindent We found that the same models that are consistent with the Galactic Center gamma ray excess, shown in ref. \cite{Caron:2015wda}, are also consistent with the small excess observed in Reticulum II, as reported by ref. \cite{Geringer-Sameth:2015lua}. The resulting $J$-factor is $\log_{10}(J(\alpha_{int}=0.5\degree)) \simeq (20.3-20.5)^{+0.2}_{-0.3}$ GeV$^2$cm$^{-5}$, which is consistent with another determination of the $J$-factor using kinematic data obtained from the Michigan/Magellan Fiber System, resulting in a $J$-factor of $\log_{10}(J(\alpha_{int}=0.5\degree)) = 19.6^{+1.0}_{-0.7}$  GeV$^2$cm$^{-5}$  as reported by  ref. \cite{Bonnivard:2015tta}. An improvement on the uncertainty of the $J$-factor will be important for interpreting these results.

\acknowledgments

We want to thank Alex Geringer-Sameth in providing the Reticulum II data and helping with problems.

\bibliographystyle{jhep} 
\bibliography{bibreport}

\end{document}